# On PQC Migration and Crypto-Agility

Alexander Wiesmaier
Hochschule Darmstadt

Nouri Alnahawi
Hochschule Darmstadt

Tobias Grasmeyer
Hochschule Darmstadt

Julian Geißler
Hochschule Darmstadt

Alexander Zeier
MTG AG

Pia Bauspieß
Hochschule Darmstadt

Andreas Heinemann
Hochschule Darmstadt

## ABSTRACT

Besides the development of PQC algorithms, the actual migration of IT systems to such new schemes has to be considered, best by utilizing or establishing crypto-agility. Much work in this respect is currently conducted all over the world, making it hard to keep track of the many individual challenges and respective solutions that have been identified. In consequence, it is difficult to judge for both individual application scenarios and on a global scale, whether all (known) challenges have been addressed respectively or what their current state is. We provide a literature survey and a snapshot of the discovered challenges and solutions categorized in different areas. We use this as starting point for a community project to keep track of the ongoing efforts and the state of the art in this field. Thereby we offer a single entry-point into the subject reflecting the current state in a timely manner.

## 1 MOTIVATION AND RESEARCH QUESTION

All cryptographic algorithms currently used are subject to security decay over time; at least due to the steady increase of computational power available to the attacker, especially considering the rise of quantum computers. As soon as sufficiently strong quantum computers exist, established asymmetric schemes like RSA, DSA, and ECDH will be broken [30, 54]. While it is not yet determined, when this will happen, there is little doubt that it will, eventually. Thus, a migration of IT-systems and infrastructures towards quantum-secure schemes has to be prepared and executed as soon as possible [86, 96].

Crypto-agility describes the feasibility of replacing and adapting cryptographic schemes in software, hardware and infrastructures. [54, 89]. It is essential in order to maintain security against future, and unknown threats [59, 89, 92]. As a negative example, in recent years, the transition from SHA-1 to SHA-256 took more than five years [59]. While the specification and implementation were done quickly, updating software and hardware products by vendors, providers, or administrators took a long time. Crypto-agile solutions should make this task easier and faster.

The question at hand is: Where are we today with regard to PQC migration and crypto-agility? This shall be answered in this paper.

## 2 SYSTEMATICS

*Research Goal and Scope:* We investigate the challenges of migrating software and IT-infrastructures to quantum-secure schemes, and introducing crypto-agility into these systems. We survey the state of the art in research and give an overview on the challenges already solved, challenges currently investigated, challenges recognized yet still unsolved, as well as current blind spots.

*Methodology:* The seed of our work is [96], which contains a variety of relevant challenges categorized into different areas, but does not provide information on their current state. We extend and refine the categorization scheme and include the respective state of the art through literature research as follows (not quite so linear):

- Initialize literature set with references of [96].
- Build keyword set from categories of [96] and "PQC", including synonyms and variations.
- Extend literature set by internet-wide literature search using combinations from keyword set.
- Work through literature, sort into category, create new category or discard.
- Extend literature set using references in and citations of categorized literature.
- Stop the process when literature set is empty.
- Put an extra effort on finding literature for empty categories.

The adapted scheme is reflected in the structure of Sections 4 and 5 and their respective overviews in Tables 1 and 2. Their content reflects the current state of PQC migration and crypto-agility. Based on this, we discuss our findings, draw our conclusions and derive future work.

*Contributions:* On PQC migration and crypto-agility, we provide:

- A literature survey of relevant papers and issues.
- An extended categorization of related issues.
- An overview of the current state of readiness of said issues.
- A discussion and pointers to open issues.
- An open community project continuing this document at https://fbi.h-da.de/cma.

## 3 RELATED WORK

This section lists work dealing with general challenges and recommendations regarding PQC and crypto-agility. The work at hand builds on that, further categorizing and discussing such challenges and recommendations, and investigating their state of readiness by correlating them with further literature that deals with individual issues in more detail.

[96] presents numerous challenges grouped into categories of PQC migration and crypto-agility. It covers a wide range of topics at a high abstraction level, yet does not necessarily offer direct information on their state of readiness. The FhG SIT[1] names twelve challenges in [93] that need to be solved to implement PQC in practical applications including migration (hybrid approaches) and agility (agile protocols and update mechanisms). [47] categorizes six main challenges for PQC in IoT and compares the performance of PQC algorithms extensively.

The Dagstuhl [2] report on "*Biggest Failures in IT Security*" [11] discusses a variety of problems in regards to achieving IT security and possible strategies to solve these. Aspects addressed include attacker models, TLS implementations and certificates. NIST[3] in [14] presents PQC adoption challenges and thoughts on migrating to PQC after the NIST standardization is concluded. Examples are a migration playbook and a way to get an inventory of used cryptography in an IT system or infrastructure.

In [24] the BSI[4] gives recommendations for action on migration to PQC and recommends the use of crypto-agile hybrid solutions and the corresponding adaptation of cryptographic protocols. [57] follow on the same footsteps with similar recommendations. They offer a brief evaluation of the current state of both post-quantum and quantum cryptography, and highlight the chances and limitations of quantum cryptography. In [26], the ETSI provides recommendations on how to make five common security standards (X.509, IKEv2, TLS 1.2, S/MIME, SSH) able to use PQC. Furthermore, they describe important use cases for cryptography and potential migration strategies to transition to post-quantum cryptography.

## 4 MIGRATION

According to [96], the migration from widely used and established cryptographic standards into post-quantum cryptography, is a transition that has to occur on numerous levels. There are basically many aspects that need to be taken into consideration in such a transition. It is thus more than a simple exchange or replacement of a cryptographic scheme or system.

[96] points out to the various contexts in which the migration will take place, and refers to many already investigated requirements and challenges, as well as new ones resulting from the characteristics of some of the new algorithms. Obviously, new cryptographic systems will need different scheme parameters, such as key, ciphertext and signature sizes. Additionally, communication and computational requirements are of equal importance. [96] also suggests that some algorithms will introduce new requirements such as state management or entropy.

In this section we highlight several aspects of the migration from conventional to post-quantum cryptography and discuss the main challenges facing this endeavor. An overview of all contributions is presented in Table 1. This includes *performance, security, implementation, process, automation & frameworks, and standards*.

---

[1] Fraunhofer Institute for Secure Information Technology
[2] Schloss Dagstuhl – Leibniz-Center for Information Technology GmbH
[3] The National Institute of Standards and Technology
[4] German Federal Office for Information Security

## 4.1 PQC Algorithms

Post-quantum cryptographic algorithms, i.e. quantum resistant, are a class of algorithms based on mathematical problems, that are solvable by neither quantum, nor classical computers; at least not under reasonable circumstances, concerning the required time and effort to solve them [17]. The first NIST report on post-quantum cryptography [30] clearly states that secure and established cryptographic schemes such as RSA, ECDSA and DSA will eventually become obsolete and no longer secure.

The current state of PQC is represented by the ongoing NIST PQC standardization process[5] which was announced [30] in April 2016. Several PQC algorithms promise to suitable replacements for the soon to be broken classic algorithms, and they fall mainly into five categories [53, 96]: Multivariate (quadratic polynomial equation), lattice-based (a grid as a discrete subset of an n-dimensional real vector space), code-based (the problem of decoding general error correcting codes), isogeny-based (algebraic geometry), and hash-based (quantum resistant hash algorithms of the SHA-family (SHA2/3)). Of the 82 submitted algorithms 69 candidates were accepted into the project and out of these, 26 advanced to the second round [5]. The evaluation criteria for submissions include: security, algorithm cost, and implementation characteristics. At most one algorithm will be selected from each underlying mathematical problem to ensure the availability of suitable alternatives considering possible advancements in cryptanalysis. Following [85], "the third-round main finalist public-key encryption and key-establishment algorithms are Classic McEliece [32], CRYSTALS-KYBER [13, 21], NTRU [29, 61], and SABER [45, 46]. Alternative candidates are BIKE [10], FrodoKEM [20, 88], HQC [82], NTRU Prime [16] and SIKE [27]. The main finalists for digital signatures are CRYSTALS-DILITHIUM [43, 44], FALCON [50], and Rainbow [39]." Alternative candidates are GeMSS [28], Picnic [52] and SPHINCS+ [18].

## 4.2 Performance Considerations

Research needs to consider the performance of PQC algorithms, because they "generally have greater computation, memory, storage, and communication requirements (e.g., larger key sizes, more complex algorithms, or both")" [96]. There are multiple research papers dealing with the performance of PQC algorithms in various facets. We classify these into the three subcategories *network performance*, *algorithm performance* and *hardware performance*.

*4.2.1 Network Performance.* Network devices have to handle the overhead introduced by PQC and still provide acceptable latency. This also impacts end user devices, e.g. smartphones, which have limited power resources and are used in mobile networks.

With larger signature and key sizes, more data has to be transferred within networks. This leads to packet fragmentation, which may lead to high delays in lossy networks [99]. In general, protocols like TLS, DTLS, IKEv2 and QUIC are able to handle bigger signature sizes, and while the overhead is significant, it is acceptable in most modern use cases [66, 106]. For example, [25] show that key establishment with Kyber-based TLS performs well in comparison to elliptic curve cipher suites. Additionally, the execution times for PQC algorithms is usually higher than for classical ones and a longer loading time for websites is a problem, because this may lead to a bad user experience [2, 76, 105]. Use cases like VPN are less impacted (establishing a connection only once) [67]. Experiments in simulated [99] as well as in real networks [77] show which algorithms are suitable for these use cases.

---

[5] https://www.nist.gov/pqcrypto

*4.2.2 Algorithm Performance.* The efficiency of PQC must allow for usage within established and new application scenarios. Notably, IoT is of special interest in terms of restricted resources.

There are various highly specific research papers dealing with this question, focusing on individual schemes. For example, [71] improve NIST round three candidate CRYSTALS-KYBER [13, 21] (KEM SHA3 and AES variant) by vectorizing time-consuming primitives. Number-theoretic transform (NTT) is another approach used to increase the performance significantly for Kyber [22], NTRU [7] and Saber [34] on Cortex-M4 chips. In general, lattice-based and code-based algorithms are more feasible than isogeny-based algorithms due to the huge performance overhead of the latter [101, 113]. Stateless hash-based symmetric signature schemes, such as SPHINCS+ [18], are generally less efficient than, e.g. lattice-based signature schemes like Dilithium [43], Falcon [49], or qTESLA [6]. They are mainly interesting for applications without strong latency requirements, such as offline code signing or certificate signing. Test results show great advantage for SPHINCS+ over similar schemes, such as Picnic, in terms of speed, signature size, and security [18]. Algorithms that require too much memory or depend on external libraries such as classic McEliece [32, 112], BIKE [10] and RAINBOW [39] pose a serious challenge on the implementation level. [15] compares round two NIST candidates and state that a higher security level increased latency and timing overhead. CRYSTALS-Dilithium [43, 44] stands out for its superior signature generation and qTesla for its signature verification. In IoT settings lattice-based schemes like New Hope [79, 80], NTRU-HRSS [61, 62] and SPHINCS+ [18] are recommended [91]. However, a decreased security level needs to be taken into account. [47] identifies post-quantum challenges for IoT devices and compares the performance of PQC algorithms and calls for more research to be done.

*4.2.3 Hardware Performance.* The performance assessment of cryptographic algorithm on specialized hardware platforms is an important decision criteria in the standardization efforts [94]. As [69] show in a benchmark for round two NIST candidates, some algorithms are still not suitable for specific types of devices and systems such as the ARM Cortex-M4 microcontroller. The deployment of algorithms with multiple parameter sets and large key sizes is still a work in progress and requires several adaptations. [15] maps high level C implementation to FPGA using High-Level Synthesis for NIST round two candidates and offers a hardware evaluation study. In general, executing cryptographic algorithms on optimized hardware leads to a significant performance boost and is essential for the usage of these algorithms, especially for less performant post-quantum algorithms in performance critical use cases [107]. In many cases, FPGAs are used to investigate the possible performance benefits [37, 72, 74].

## 4.3 Security Considerations

The new post-quantum algorithms introduce additional mathematical concepts into the cryptographic world. New security risks that may come with these concepts have to be investigated. Moreover, the choice of the algorithm parameters and key sizes play a vital role in establishing security of a cryptographic scheme. And last but not least, the security of a scheme can often be threatened by side channel attacks based on timing or power consumption. We divide the security issues into three subcategories, those being *algorithm/parameter selection*, *cryptanalysis* and *side channel attacks*.

*4.3.1 Algorithm/Parameter Selection and Trade-offs.* Due to the described performance challenges that come with the new post-quantum algorithms, depending on the use case a different algorithms and/or set of parameters may be suitable. It is important to understand the trade-off between security and algorithm requirements to be able to choose the right algorithms and/or parameters for different use cases.

The most common difficulties arise from the integration of PQC algorithms into existing protocols. While the implementation and performance of those algorithms itself doesn't seem to add unfeasible overhead to most applications, most protocols impose tight restrictions that cannot handle the much bigger key and/or signature sizes. [67, 101].

Algorithm selection does greatly depend on other sections and future developments. As of now, it seems that no perfect algorithm will be found that can replace existing systems and provide the same security levels with the same performance that we are used to today. If this remains the case, carefully selecting algorithms for specific use cases will be an important topic in the future.

*4.3.2 Cryptanalysis.* Cryptanalysis examines the security of cryptographic schemes given a sophisticated attacker that is able to perform analytical and/or practical attacks. Analytic attacks focus the mathematical basis and the overall scheme whereas practical attacks use statistical methods to gather information about the key or plain text.

Many of the candidates that were submitted to the first round of the NIST challenge have fallen to various cryptanalysis attacks. In general, either the imposed security of the attacked algorithms fell under the requirements of NIST [38], or the scheme was completely broken and the attackers were able to retrieve the private key [104]. Some schemes seem to be secure under classical computing preliminaries only, but fail to resists quantum attacks [41]. On the other hand researcher extracted indications for secure designs from those attacks. [102] present a method to assure security against a certain type of attacker and [64] present "models that enable direct comparisons between classical and quantum algorithms", improving security assumptions for PQC algorithms. In addition PQC can not only be used for key exchange and signatures, other areas can be pseudo-random number generators, ultimately supporting cryptography in other ways than direct encryption [75].

*4.3.3 Side Channel Attacks.* These attacks target the specific circumstances of algorithm execution like timing and power consumption. PQC algorithms introduce new attack surfaces in this regard and have to be analyzed for weaknesses not only in a general regard of the algorithm but also in specific hardware circumstances.

For all categories of PQC algorithms, side-channel attacks were found, and in many cases countermeasures were proposed. In [33], an overview of several PQC-related side-channel attacks is given. It discusses side-channel attacks for the different categories of PQ algorithms, random number generators and physically unclonable functions. The paper states that side-channel attacks on PQ algorithms and their countermeasures are still in a very early state. Many side-channel attacks on PQC are not yet evaluated and the current countermeasures usually use ad-hoc designs that protect against individual side-channel attacks, but remain vulnerable against others.

Other papers show that many attack vectors have been found, and at the same time effective countermeasures for these attacks have been introduced [117, 118], while there is also an effort to minimize attack surface overall [70, 109, 118]. However, such attacks are not restricted to the new PQC algorithms, as they can still target some of the standard building blocks in many of the old and new schemes alike. As [68] indicates in their single trace soft-analytical side-channel-attack on KECCAK, that it was possible to exploit some vulnerabilities on 8-bit microcontrollers. Countermeasures such as masking and hiding are therefore recommended.

[73] introduce potentially harmful attack vectors but only partially succeed in recovering the key, whereas [40] introduce the "singularity attack", which successfully breaks signatures of the multivarite public key scheme Himq-3 by using a system of linear equations in which the solutions leak parts of the private key.

## 4.4 Algorithm Migration Process

There are many questions regarding possible approaches to the migration process, depending on the use cases and the protocols at hand. Moving a set of software applications, their implemented algorithms or changing their underlying infrastructure requires a tedious process, which involves thorough planning and execution over a period of time. [96] indicates that the migration process is not simply limited to algorithm exchange or replacement. Therefore, one cannot assume a successful migration without considering the challenges related to the planned approach. This is especially challenging when the systems to be migrated need to stay online and/or must maintain interoperability with other systems that may or may not (yet) be upgraded.

There are multiple considered or planned approaches regarding the PQC migration on the algorithm level. These approaches address different issues and pose questions regarding backward compatibility, the definition of hybrid key exchange protocols, and message and key sizes depending on the chosen scenario.

*4.4.1 Hybrid.* "The primary goal of a hybrid mode is to ensure that the desired security property holds as long as one of the component schemes remains unbroken." [36] This approach uses two or more independent algorithms chosen from both post-quantum, and classical schemes. This way, it is not necessary to fully trust either algorithm. On the one hand, the relatively new post-quantum scheme until it had proven secure enough. On the other hand the potentially outdated classic scheme.

Although many of the NIST candidates — such as BIKE [10], SIKE [27] and NewHope [8] — prove suitable for usage in a hybrid scheme, some protocols have size constraints that prevent some schemes from being used. A trade-off between increasing the size limits and performance needs further investigation for more precise results. First implementations and case studies [36, 116] using TLS 1.2, TLS 1.3 and SSH in openSSL reveal some challenges regarding message size for the protocol itself. [36] suggest two options for conveying the cryptographic data for multiple algorithms. Either by extending the format of the protocol message, or through concatenating the data into a single value. The method of combining also effects the security and performance of the protocol. This can be done by concatenating or XORing. Their experiment on key exchange using TLS 1.2 shows that adding PQC algorithms to the original cipher suite could lead to *combinatorial explosion*, since the cipher already contains the key exchange and authentication methods, the symmetric cipher and the hash function.

Google experimented [23] with a hybrid approach for certain connections to its servers. The solution utilizes the standard use of elliptic curves with lattice-based New-Hope. The findings of the experiment raise awareness for lattice-based algorithms; and for testing the feasibility of post-quantum cryptography, which resulted in no network problems and a median connection latency increase of one millisecond [1]. Another experiment [77] compares three groups using post-quantum CECPQ2, CECPQ2b or non-post-quantum X25519. In most cases, CECPQ2 performed better than CECPQ2b due to its smaller computational costs.

The use of lattice-based PQC algorithms with the Apache Kafka Software using a hybrid approach [113] shows that the performance of the hybrid solution is as good as the slowest algorithm used within. [101] examine the use of lattice-based PQC algorithms for the industrial protocol Open Platform Communications Unified Architecture. The approaches are feasible, but result in additional performance and communication overhead.

*4.4.2 Combiner.* In contrast to hybrid schemes, a combiner takes multiple cryptographic components and combines them into a new one. These are secure, as long as one of the original components is secure. Combiners can be used to create hybrid KEMs and thus key exchange protocols.

[19] present "new models for KEMs and authenticated key exchange protocols that account for adversaries with different quantum capabilities". They also propose several combiners and a "provably sound design for hybrid key exchange using KEMs." [48] state a "notion for robust multi-property combiners" and implement a combiner that satisfies the strongest notion by preserving every property of the input hash functions. They "concentrate on the most common properties [...], which are collision-resistance (CR), target collision-resistance (TCR), pseudorandomness (PRF), message authentication (MAC), one-wayness (OW) and indifferentiability from a random oracle (IRO)."

*4.4.3 Composite.* Another approach to combine cryptographic data is possible through the usage of composite public keys and signatures. Unlike the hybrid approach, this method uses multiple post-quantum algorithms in one uniform format to produce a composite key structure.

[97] propose a draft on composite keys and signatures for PKIs. Their proposal encodes a sequence of public keys and signatures so that they fit in the existing public key and signature keys in PKCS#10, CMP, X.509, and CMS structures. They provide a mechanism to address algorithm strength uncertainty by encoding multiple PKs and signatures into existing ones, as well as an algorithm for validating.

## 4.5 Automation and Frameworks

Even though the presented approaches to migration may well solve problems concerning algorithm implementations and protocol definitions, a full scale migration of IT-systems and infrastructures involves more complex issues. [96] raises many questions regarding the possible scenarios and context of the migration process on a higher level of abstraction. First, one has to identify the domains and their respective priorities in the planned migration. Moreover, there are interrelated dependencies that may not be easy to resolve. Thus, the migration could prove difficult in IT-infrastructures with security vulnerabilities, making them prone to failures.

RFC6916 formalizes the migration process for algorithm suites in the Resource Public Key Infrastructure (RPKI) [51]. While the process of a *key-rollover* is already implemented and in use, this RFC targets the migration of an algorithm suite to another. The process defines five distinct milestones that slowly enables the usage of the new algorithm and at the same time inverse proportionally limits the usage of the old one until it is deprecated and not used anymore.

[42] propose a framework for the security analysis of hybrid authenticated key exchange protocols and introduce the Muckle protocol. Muckle combines QKD, classical and post-quantum components and promises security against a broad class of adversaries, thus making it a good candidate to build more complex protocols like TLS onto it. A challenge remains in finding a suitable QKD to support this protocol.

## 4.6 Towards New Standards

Awaiting the third NIST PQC conference in June 2021, several post-quantum cryptographic algorithms have been continuously analyzed and benchmarked. Their implementations are tested in applications and in standard cryptographic and communication protocols. As the NIST standardization process is still ongoing [85], many candidates have already proved a suitable replacement for different classical schemes and showed capable of integration within the widely used cryptographic protocols and applications utilizing them.

"Trade-offs between computational efficiency and communication sizes" [106] seem to be the first major concern in standardizing cryptographic primitives. [106] present the Open Quantum Project, which consists of the libqos library. The library provides quantum resistant cryptographic primitives, integrated within cryptographic applications like OpenSSL. They compare NIST Round 2 PQC candidate implementations using OpenSSL. The implementation shows that the differences in performance in the standalone comparison have less impact on the overall performance because of network and additional overhead. This research supports the general efforts in establishing new standard cryptographic schemes.

[35] propose an RFC draft for hybrid key exchange schemes for TLS 1.2 combining "ECDH with a post-quantum key encapsulation method (PQ KEM) using the existing TLS PRF". They utilize the "Bit Flipping Key Exchange (BIKE) [10] and Supersingular Isogeny Key Exchange (SIKE) [27] schemes in combination with ECDHE as a hybrid key agreement in a TLS 1.2 handshake."

[87] evaluate three PQC algorithms from the third round of the NIST PQC standardization process, that are suitable for DNSSEC within certain constraints. Although they meet the most important requirement for the signature size, they require much larger signature and public key sizes compared to classical algorithms. None of them displays a perfect alternative, so changes to the DNSSEC protocol are likely needed. Other possible solutions could include increase of support for TCP in general or DNS-over-TLS or DNS-over-HTTPS. Transitioning should begin early and hybrid technologies can help to achieve it.

## 4.7 Open Issues

Our survey reveals a major focus of the community on how PQC schemes can be implemented and combined, as well as their performance and security. Moreover, considerable work on the integration into established standards such as TLS, SSH, X.509, etc. has been done. In this section, we address open questions regarding the PQC migration.

*4.7.1 PQC Algorithms.* The NIST PQC standardization is ongoing. Given the rigorous selection process with worldwide participation and a current set of 15 finalists, there should be little doubt that we will soon have a set of vetted PQC algorithms at our disposal. Still, there is enough room for further development to either adapt to special requirements (cf. 5.6.5) or to meet future developments.

*4.7.2 Performance Considerations.* The examples in 4.2 show that, depending on the choice of algorithm and field, PQC can be very well a replacement for classic algorithms. However, it is not very clear, which algorithms perform best in which situation, moreover on which hardware platforms or for which application scenarios. For example, larger key sizes lead to higher loading times, negatively impacting normal user experience in web related use cases. This also applies to IoT, which is still an underrepresented field where more research needs to be done [47].

*4.7.3 Security Considerations.* Since some PQC algorithm aspects are still fairly new and didn't face enough scrutiny, the long term security of these algorithms remain questionable. Especially attack vectors like side channel attacks are still in an early state. A very special technique that has not gotten enough research so far is the exploitation of location-based leakages. [9] exploit the spatial dependencies, such as the dependency of the secret key and certain registers, of cryptographic algorithms with the help of near-field micro-probes that are able to measure small regions on a chip with a very high resolution. Many vulnerabilities have been found and fixed, but a crucial question remains, if less costly mitigation measures can be implemented that impact the overall performance as little as possible.

*4.7.4 Implementation Considerations.* Implementations of complex mathematical models are challenging tasks that may introduce new vulnerabilities that were not present in the mathematical models themselves. Complexity as a whole can introduce many bugs and has to be managed and kept down as much as possible. Code quality and maintainability is a crucial design aspect that is often neglected in cryptographic code. The research community should take into consideration, that the code not only needs to be correct, but can also be maintained by developers and understood by auditors. Best practices from decades of software engineering have to be taken into account to guarantee high code quality.

**Table 1: Overview Migration**

| | |
|---|---|
| **PQC Algorithms** | |
| NIST PQC selection | [5, 30, 85] |
| **Network Performance** | |
| High delay in lossy networks | [99] |
| TLS, DTLS, IKEv2, QUIC | [25, 66, 106] |
| Bad user experience | [2, 76, 105] |
| VPN procotol evaluations | [67, 77, 99] |
| **Algorithm Performance** | |
| Improve CRYSTALS-KYBER | [22, 71] |
| Lattice vs. isogeny | [101, 113] |
| PQC for IoT | [47] |
| Lattice-based cryptography as alternative | [79, 80] |
| NIST PQC candidates | [8, 10, 13, 16, 18, 21, 27–29, 32, 39, 44, 45, 50, 62, 65, 82, 88] |
| **Hardware Performance** | |
| CRYSTALS-Dilithium, qTesla | [15] |
| Performance critial use cases | [107] |
| FPGA performance benefits | [37, 72, 74] |
| **Algorithm & Parameter Selection** | |
| Key/sig. size problematic for protocols | [67, 101] |
| **Cryptanalysis** | |
| PQC schemes broken by cryptanalysis | [38, 41, 104] |
| New security assessment methods | [64, 102] |
| Code-based PQC algorithms for PRNG | [75] |
| **Side Channel Attacks** | |
| Side-channel attacks | [33] |
| Minimize attack vectors | [70, 109, 118] |
| Successfull attack on Himq-3 | [40] |
| **Algorithm Migration Process** | |
| Hybrid TLS & SSH Impl. | [36, 116] |
| Hybrid Lattice-Based | [1, 23, 101, 113] |
| Hybrid PQ CECPQ2(b) & X25519 | [77] |
| **Automation & Frameworks** | |
| RFC6916 PKIs Process Formalization | [51] |
| Muckle Protocol Security Analysis | [42] |
| **Towards New Standards** | |
| NIST Report on Round 3 Finalists | [85] |
| Review of NIST Candidates | [56] |
| Open Quantum Project | [106] |
| DNSSEC PQC Draft | [87] |
| Decentralized Cert. Management | [55] |

*4.7.5 Algorithm Migration Process.* The methods introduced in 4.4 address issues on the implementation and communication levels, and offer promising solutions for many challenges; such as defining hybrid formats, algorithm negotiation and parameters. However, some of the biggest issues regarding this approach are related to design aspects. For example, how many algorithms to combine within a hybrid protocol, and how to do so. Moreover, negotiation strategies need to be determined for all possible scenarios depending on the communicating systems [36]. At this point, the problems of dealing with legacy systems and deprecated algorithms are still left out as well and require further research and more concrete solutions.

*Algorithm Negotiation and Backward Compatibility.* Cipher-suite negotiation determines an algorithm that all parties are willing to use. Downgrade attacks target these negotiations to force a weaker security. Depending on the scenario, different design decisions are required, especially considering the negotiation strategy. [36] distinguish between three main cases, where the client, the server, or both are either aware or unaware of the new protocols. This implies a possible downgrade on the side that is already using a new protocol; at least when not using TLS v1.3.

*Legacy Systems.* Since not all systems will be updated to post-quantum algorithms in time or cannot be updated at all, strategies on how to deal with these systems have to be developed. These could be automation frameworks, which are used transparently within protocols or systems to manage their migration into PQC, or ensure their security.

*4.7.6 Automation and Frameworks.* [96] emphasizes the need for automated software in order to deploy PQC algorithms and protocols with minimum human interaction; especially in data centers and cloud-based applications. Such software should be able to identify the currently used cryptographic components such as public key schemes, algorithm versions, ports and devices. Automated tools could also support real-time analysis of an ongoing migration, check for any weaknesses and verify the security of new PQC libraries. The aforementioned issues clearly point out the importance of developing frameworks capable of managing, executing and testing the various aspects within a migration process on different scales and implementation levels. Our survey shows a lack of such automated software. Too few projects and research dealing with the aforesaid questions were found. Most of the offered solutions address relevant issues, yet rather on a low level of abstraction, and focus on algorithm and / or protocol design, integration and communication.

*4.7.7 Towards New Standards.* The efforts undertaken towards establishing new cryptographic standards are promising, yet still focus more on low level cartographic primitives and schemes. For algorithms, there is one formal standardization process (by NIST) focusing on their mathematical foundations and implementation characteristics, as shown in section 4.6. The efforts made for testing, bench-marking or integration in new or existing protocols consist mainly of independent research and draft proposals, such as the hybrid TLS 1.2 RFC draft [35] and RFC6916 for Public Key Infrastructures (RPKI) [51], which are not sufficient for the overall migration process on a broader scale, at least not yet. The discussed processes however, seem to suffer from lack of interest in the strategies required to execute the actual migration on a more sophisticated level of abstraction.

## 5 AGILITY

Given the already complex problems that migration to a post-quantum future presents, there is an even more general aspect of the problem of cryptography that has so far been unaddressed by the community. As quantum computing is still a relatively young research field and IT security is in constant flow anyway, there is a high chance that future changes need to be made to cryptographic primitives and protocols. Regarding these preliminaries cryptographic algorithms face the challenge of designing in the ability to adapt to those changes in a way that exchanging core cryptographic primitives and routines should be practically feasible and simple. [96] therefore proposes a "new challenge" of cryptographic agility that should be researched and investigated in. Cryptographic agility (crypto-agility for short) should be clear in its effect and security impact and also be applied with a feasible overhead to any kind of computational domain, including but not limited to algorithms, protocols, applications and even distributed systems and infrastructures.

In this section we present aspects regarding the notion of crypto-agility and offer a brief discussion of the open issues. This includes *modalities, development considerations, testing, incentives, and frontiers of cryptography.* An overview of all contributions is presented in Table 2.

### 5.1 Modalities

Beside algorithmic agility, [96] proposes several other aspects, where different forms of crypto-agility can be manifested. These supposed modalities display agility in a broader view of the topic. The core characteristic of agility and also of those modalities, is the ability to adjust to a certain environment without any or with as little as possible human intervention. To be able to do so, there has to be either a certain configuration, policy or other threshold that is present before the algorithm starts or the algorithm has to be "smart" enough to figure out its application scope and environment itself and adapt accordingly.

*Implementation Agility* tries to shed light onto the question how the process of implementation can be supported so that algorithms or applications can be crypto-agile. A common first approach is to propose new [110, 111] or enhance existing protocols like TLS [98], PKINIT [12] or APIs [98] to be able to adapt to new cryptographic schemes and negotiate their usage with other parties. Other approaches expand existing infrastructure to be able to exchange cryptographic algorithms [31, 81, 108].

Another modality is the ability to compose cryptographic keys or signatures from other signatures and keys called *Composability Agility*. [97] are working on a draft for composite keys and signatures for PKIs so that a sequence of public keys and signatures can fit in the dedicated spots for simple public keys and signature in various cryptographic structures.

### 5.2 Development Considerations

Crypto-agile libraries and APIs should minimize the manual maintenance of any developer adjusting source code when new PQC algorithms need to be used. In an ideal world, a developer would only update the libraries version and nothing else. The idea is to have widely used cryptographic libraries with an crypto-agile design for example similar to the eUCRITE API [114, 115], where templates for PQC operations are used while the method calls stay the same. An update mechanism is currently being researched and developed. Also [100] advocates for user-friendly cryptographic APIs with focus on crypto-agility and secure update mechanisms.

Already used frameworks with algorithmic crypto-agility in mind are JCA/JCE, TLS, SSH and X.509 v3 digital certificates. Both TLS and SSH utilize a cipher negotiation mechanism on a communication level. X.509 certificates added flexible extensions in version 3, where in an hybrid cryptography approach a PQC public key and a signature can be embedded next to the conventional data. [96]

Current research is thin in terms of new approaches providing crypto-agility and widely distributed between fields. [110] propose a negotiation protocol similar to TLS but layer independent. In an API setting [78] implement a plug-in structure in the *Crytography API: Next Generation* from Microsoft to exchange cryptographic algorithms without any change to the code of the program. Crypto-agility can also be established through design, as [108] show that key length and hash and signature algorithms can be adapted. [58] follow the same design principle in being algorithm independent.

Next generation libraries for quantum-resistant cryptography like liboqs [6] of the Open Quantum Safe [7] project are great initiatives getting new prototypes developed in an open source matter, but are missing a crucial design principle, being crypto-agile.

In general building crypto-agile schemes will introduce new complexity and create new attack surfaces. The question is what new attack surfaces will arise [96]. Until more generalized schemes are designed, widely used and tested, understanding attack surfaces and their influence on the design will be an accompanying research area. Today's research is mostly based on already known schemes like negotiation-based protocols.

Another research area focuses on understanding the usability of cryptographic API, where [4] examine the usability of cryptographic libraries and found poor documentation and missing code examples to be an issue. They call for simple interfaces and code examples for common tasks. As a solution [83] present a web platform for cryptographic code examples with an experiment where the participants were more effective in solving the task and the code more secure. In [84] the same authors found that code examples need to be more concrete to have a meaningful impact. Similarly [60] examine a prototypical implementation of a documentation system for two crypto APIs in an experiment with 22 probands.

### 5.3 Testing

[96] notes several aspects of cryptographic agility that need more research but also due to their nature build on top of other aspects that have to be better understood first.

---

[6]https://github.com/open-quantum-safe/liboqs
[7]https://openquantumsafe.org/

Careful and thorough testing and validation as one of these aspects should support all steps of cryptographic processes from implementation to roll out on greater infrastructure to updating minor components [96]. Developments and lessons learned from the area of software engineering to compliance policies that enforce testing of components can supplement this process. Effective approaches will have to consider "relations based on the specifications of the algorithms, and design test cases such that a systematic coverage of the input space is achieved." An important aspect here is also the quantification of test coverage. [103].

## 5.4 Incentives & Best Practices

Following [96] it is important to provide proper incentives to implement agility into software. [11] propose an app-store-like ranking system, where adherence to security best practices results in higher rankings, as incentive mechanism. In the category current best practices, RFC 7696 [59] provides a guideline on crypto-agility and algorithm selection for IETF protocols. There, advantages of specific design decisions are discussed and considerations for individual implementations are introduced, often resulting in trade-offs between security on one hand and usability, interoperability or agility on the other. Recognizing that crypto-agility presents a variety of challenges, from hardware issues to processes and organizational aspects within enterprises and entire industries, [100] identified important building blocks towards crypto-agility, namely, APIs, update mechanisms, and documentation.

## 5.5 Frontiers of Cryptography

At the end of this section we are left with the question on how certain categories of cryptography like fully homomorphic encryption (FHE), password-authenticated key agreement (PAKE), blockchain or threshold cryptography are going to react to migration and cryptographic agility. Some of these put constraints on the amount of agility they are able to support. Blockchain technologies for example seem prepared for PQC but cannot be agile because of their protocol [95]. Other areas that face heavy constraints to their ability to implement agile solutions are satellite based communication systems. These systems can adapt to PQC but can only be updated in very limited circumstances if at all [90]. The notion of agility and its applicability to cryptographic primitives is addressed in [3]. They provide a formal analysis of different primitives showing when they are considered agile, and when not. One core idea is that some primitives are not agile in their nature, but could be if combined with others forming a set of schemes.

## 5.6 Open Issues

The previous sections presented crypto-agility in the light of various modalities and also highlighting development considerations, testing and in addition to that, sections that deal with more general issues of motivation and contradictions. Closing this section we again regard the unanswered questions and try to identify areas that the research community has not treated with enough focus.

*5.6.1 Modalities.* Crypto-agility can be viewed as a generalization of migration, as it formalizes the ability to permanently change. Connected with this generalization is the wish to make migration as a whole more approachable and simpler. This can be seen from many different angles, which are connected to each other. Modalities like implementation agility can be solved through frameworks. However, ensuring the desired cryptographic agility within the relevant IT systems cannot be considered a practical endeavor. Currently there are neither automated tools, nor dedicated frameworks capable of executing and validating a cryptographic migration process; let alone sustaining any cryptographic agility after the expected migration. Such automation should preferably be able to identify, analyze, deploy or replace cryptographic components within IT systems; following well studied and planned strategies and guidelines. Composability agility on the other hand is the focus of several research groups through their work of combining and composing sets of keys and signatures for cryptographic schemes.

Other modalities did not get enough attention so far, which is partly due to the fact that they will naturally be investigated at a later stage of research. Overall most modalities can be summarized under the general term of context agility, the ability to adapt to a different context. And while as a future goal this ability seems to be a promising research candidate that might solve many current issues, it is still a long way to reach it.

It should be mentioned that all of these abilities and modalities could introduce severe vulnerabilities but at least will introduce more complexity which should only carefully be increased.

*5.6.2 Development Considerations.* 5.2 shows that we still need cryptographic libraries following crypto-agile design principles. Current approaches are cipher negotiations or utilizing hybrid cryptography. Research for different architectural ideas is thin and widely distributed between fields and mainly concentrates on improving cipher negotiation schemes, working with plug-in structures and adaptable key lengths hash and signature algorithm. We need a change in direction for developing current or new cryptography libraries. If these don't follow crypto-agile principles it will bring us back to the issue we have now, that constant adjustment are needed and development/maintenance costs will occur. Nonetheless, new crypto-agile schemes will introduce new complexity and therefore attack surfaces, that need to be researched and understood. In terms of user interfaces for different end users, questions asked in [96] are still unanswered. Until software development principles don't change and other general purpose crypto-agile schemes are presented, questions about attack surfaces and user interfaces will be on hold and will play a role in the future again.

**Table 2: Overview Agility**

| Modalities | |
|---|---|
| New agile protocol | [110, 111] |
| Enhance existing protocols for use with PQC | [12, 98] |
| Enhance existing infrastructure for PQC | [31, 81, 108] |
| Draft for composite keys and signatures | [97] |
| Development Considerations | |
| eUCRITE API | [114, 115] |
| Research on CA mechanism | [78, 100, 110] |
| CA as design principle | [58, 108] |
| Eval. crypto libs | [4] |
| Eval. code examples for crypto libs | [83, 84] |
| Eval. docum. system for crypto libs | [60] |
| Testing | |
| Algorithm relations for better test coverage | [103] |
| Incentives & Best Practices | |
| Ranking by best practice as incentive | [11] |
| Best practice for agility in protocols | [59] |
| Building blocks of crypto-agility | [100] |
| Frontiers of Cryptography | |
| Blockchains difficult | [95] |
| Satellites difficult | [90] |
| Cryptographic primitives handable | [3] |

*5.6.3 Testing.* To ensure crypto-agile software is used correctly and enables the advantages of cryptographically secured communication, it has to be checked, tested and validated. Testing and validation can happen on many levels, from checking the mathematical models, over source code audits, up to checking the entire infrastructure whether every component works in its assigned role. As research in the crypto-agility field is still in an early stage, testing and validation has to be examined more.

*5.6.4 Incentives & Best Practices.* If looking at the effort and time it took for historic cryptographic updates (e.g. DES, RC4, MD5 or SHA-1) to be executed is not incentive enough, the nowadays very strict data protection laws in many countries around the world (e.g. GDPR in Europe) along with their respective legal (i.e. monetary) consequences should be sufficiently convincing. The question which incentives for introducing crypto-agility could or should be given is still open. Besides the possible competitive advantage mentioned in 5.4, further answerers could include adherence to existing laws (e.g. GDPR in Europe) and reduction of cost (e.g. for switching algorithms). It is also thinkable that new regulations will appear, demanding crypto-agility for applications in certain contexts, e.g. for critical infrastructures. All said apart, we need to find out how to motivate the developers to want to weave crypto-agility into their code. Equipping the very crypto libraries with (semi-)automatic agile features will surely lower the bound. Regarding best practices, RFC 7696 [59] (cf. 5.4) only covers a small fraction of the vast field of possible application areas of crypto-agility. Further research and hands-on development is needed here, too. The community must develop a common understanding including definitions, terms and language in general in order to support the discussion on crypto-agility and place the various contributions on a common ground. In our view, a widely accepted reference model should be developed for this purpose. As seen at the example of the ISO/OSI network reference model [63], this can prove very useful even if not implemented.

*5.6.5 Frontiers of Cryptography.* Given the task of constant change, the question is left whether emerging cryptographic schemes and applications, that can differ considerably in structure and use case from current ones, are able to adapt to migration and crypto-agility. The research community has to make sure that these new categories of cryptography, such as fully homomorphic encryption, password-authenticated key agreement, blockchain or threshold cryptography, are examined. Some cryptographic applications like Bitcoin have to come up with new solutions regarding the permanent danger from broken cryptographic algorithms.

# 6 CONCLUSION

Considering the expected rise of quantum computing, challenges posed upon classical cryptography and the IT systems using them will become tangible threats jeopardizing the safety and security of IT systems, applications and communication alike. New cryptographic algorithms, such as NTRU, Rainbow, and BIKE, cannot yet be considered fully understood, nor fully secure, as they still require further investigation, testing, formal modeling and analysis. Additionally, finding new suitable standards instead of the soon out-dated cryptographic schemes and developing new standards are only the first steps. Moreover, a full scale migration of cryptographic systems, including the IT systems and infrastructures using them, is a task that cannot and should not be taken lightly. It is a long ongoing process that requires thorough planning and execution on many different levels.

In this paper, we present a survey of the ongoing research efforts towards realizing a migration from classical to PQ cryptography, as well as achieving crypto-agility within systems relying on cryptographic schemes for their security. Our findings indicate a major focus of the ongoing research on algorithms per se, and the feasibility of migrating different cryptographic schemes and security protocols with reasonable trade-offs regarding algorithm performance and communication efficiency. However, there are still many challenges and open issues that need to be addressed, especially on higher levels of abstraction. This includes protocol design, migration and deployment strategies, legacy systems, testing frameworks and process automation. One of the most important challenges to consider is creating crypto-agile systems capable of handling the diversity accompanying the parallel usage of old and new schemes during the migration phase. Moreover, sustaining crypto-agility after the migration and maintaining the same level of safety and security, is also of great importance. Last but not least a common view and understanding on crypto-agility, for example in terms of a reference model, is missing.

# 7 OUTLOOK

Identifying the challenges on the way towards usable and agile PQC is the first step of our research. The overview and findings at hand provide a starting point for providing answers to the yet untouched issues. Theoretical and practical case studies can be conducted to verify the proposed challenges and maybe find new ones in real-world scenarios, and come up with new ideas.

We initiated a community project on https://fbi.h-da.de/cma, allowing participants to continue the work started with the paper at hand and share their findings and ideas. At this point, our research group is also working on the development of a cryptographic solution directed towards aiding the community efforts in achieving crypto-agility. A first prototype of a cryptographic interface called eUCRITE[8]; providing both conventional and PQC functionalities is developed in the form of a general purpose API for the Java programming language in an effort to enable usable crypto-agility for high level developers and software designers. This API is currently still under development, and is currently being extended with an automated update mechanism to ensure constant usage of the latest secure cryptographic schemes, without interrupting the system operations. We have plans to develop an automated cryptographic infrastructure detection tool capable of identifying the cryptographic components used within a given IT system. Such a tool can prove very valuable in overcoming the complexity of the otherwise manual process of finding all the interconnected cryptographic measures. This tool will then be tested and evaluated in real-world settings in a real world IT system.

---

[8]https://fbi.h-da.de/eucrite

*Acknowledgment:* This research work has been funded by the German Federal Ministry of Education and Research and the Hessian Ministry of Higher Education, Research, Science and the Arts within their joint support of the National Research Center for Applied Cyber-Security ATHENE.